\documentclass[twocolumn]{revtex4}

\usepackage{soul}
\usepackage{amsmath}   
\usepackage{amssymb}
\usepackage{graphicx}
\usepackage{dcolumn}
\usepackage{bm}
\usepackage{amsmath}
\usepackage{graphicx}
\usepackage{amsfonts}
\usepackage{subfigure}
\usepackage{graphicx}
\usepackage{array}
\usepackage{float}
\usepackage{color}
\usepackage{amssymb}%
\usepackage[colorlinks=true,linkcolor=blue]{hyperref}%
\hypersetup{allcolors=blue}
\usepackage[normalem]{ulem}
\usepackage{xcolor}

\begin{document}

\title{Measurement of Rb $g$-series quantum defect using two-photon microwave spectroscopy}
\date{\today }
\author{K. Moore}
    \altaffiliation[Present address: ]{SRI International, Princeton, NJ 08540, USA.}
\affiliation{Applied Physics Program, University of Michigan, Ann Arbor, MI 48109, USA}

\author{A. Duspayev} 
 \email{alisherd@umich.edu}
\author{R. Cardman}
\author{G. Raithel}
\affiliation{Department of Physics, University of Michigan, Ann Arbor, MI 48109, USA}

\begin{abstract}
We utilize two-photon high-precision microwave spectroscopy of $ng\rightarrow(n+2)g$ transitions to precisely measure the high-angular-momentum $g$-series quantum defect of $^{85}$Rb. Samples of cold Rydberg atoms in the $ng$ state are prepared via a three-photon optical excitation combined with controlled electric-field mixing and probed with 40-$\mu$s-long microwave interaction pulses. The leading systematic uncertainty arises from DC Stark shifts, which is addressed by a cancellation of background electric fields in all three dimensions. From our measurements and an analysis of systematic uncertainties from DC and AC Stark shifts, van der Waals interactions, and microwave frequency calibration, we obtain $\delta_0=0.0039990(21)$ and $\delta_2=-0.0202(21)$. We discuss our results in context with recent work elsewhere, as well as applications towards precision measurement.
\end{abstract}

\maketitle
\section{Introduction}
\label{sec:intro}

Measurements of atomic transition frequencies are the cornerstone of precision spectroscopy, used in applications ranging from atomic clocks \cite{bloom, Martin, Milner} to measuring gravitational redshifts \cite{chou} and the radius of the proton \cite{pohl, Beyer, Fleurbaey, Bezginov}. Often, cold atoms are used in these measurements. Alkali atoms, which have a single valence electron similar to hydrogen, are easier to laser-cool than hydrogen due to a lower recoil energy and near-infrared cooling-transition wavelengths. However, in an alkali atom such as rubidium, the interaction between the ionic core of the atom and the valence electron depresses the energy levels of the valence electron below the expected hydrogenic levels (the``quantum defect"). In precision spectroscopy, it is imperative to determine this quantum defect for each commonly-used alkali species. Moreover, precision measurements of quantum defects can serve as a check for advanced theoretical calculations and contribute to a better understanding of the electronic structure in atoms. \par
Here, we measure the $ng$-series quantum defect of Rb ($n$ is the principal quantum number). For electrons in high-angular-momentum states, the quantum defect is dominated by the polarizability of the ionic core, which can be extracted from quantum defect measurements. In the most recent experimental measurements of the $ng$-series quantum defect, microwave spectroscopy of $nd\rightarrow(n+1)g$ transitions~\cite{lee} and $nf\rightarrow ng$ were performed~\cite{berl}. In our work, we use sub-THz spectroscopy to measure $ng\rightarrow(n+2)g$ transitions in an environment with three-dimensional electric field control in the atom-field interaction region. Our two-photon transition depends only on one set of quantum defects and takes advantage of equal Land\'e-$g$ factors in the lower and upper states, making the measurement insensitive to external magnetic fields. We measure the $ng$-series $\delta_0$ and $\delta_2$ quantum defects (where $\delta_0$ and $\delta_2$ are the Ritz expansion coefficients \cite{drake}) of $^{85}$Rb with a precision comparable with recent experiments~\cite{lee, berl}. Our results agree with~\cite{lee} within the stated uncertainties, but are slightly outside of the uncertainty overlap with~\cite{berl}. \par

This field of research leads to a better characterization of hydrogen-like species for future precision measurements. Specifically, the results may pave the way towards an improved Rb$^+$ polarizability measurement, which is necessary for precision measurement of the Rydberg constant using circular Rydberg states of nonhydrogenic atoms, which are easier to laser-cool and to excite than hydrogen~\cite{haroche, tan, ramos}. A Rydberg-constant, $R_{\infty}$, measurement can help to solve the proton radius puzzle, for which an inaccurate value of $R_{\infty}$  has been named as a possible answer~\cite{pohl, Beyer, Fleurbaey, Bezginov, pohlscience}.  

\maketitle
\section{Methods}
\label{sec:methods}

The objective of this study is to obtain values for the $ng$-series $\delta_0$ and $\delta_2$ quantum defects in the Rydberg-Ritz formula,

\begin{equation}
\label{eq:1}
\delta(n) = \delta_0 +  \frac{\delta_2}{(n-\delta_0)^2} + ... \quad,
\end{equation}

\noindent from spectroscopic measurements of transition frequencies $\nu_{n_1, n_2}$ between $g$ Rydberg states with principal quantum numbers $n_1$ and $n_2$,

\begin{equation}
\label{eq:2}
\nu_{n_1, n_2} = R_{Rb} \ c \ \left( \frac{1}{(n_1 - \delta(n_1))^2} - \frac{1}{(n_2 - \delta(n_2))^2} \right). 
\end{equation}

\noindent Here, $R_{Rb}$ is the Rydberg constant for $^{85}$Rb. To improve precision, several redundant measurements are performed. The accuracy of the measurement depends critically on our analysis of systematic shifts in Sec.~\ref{sec:unc}. \par

\begin{figure}[t]
 \centering
  \includegraphics[trim=15 0 0 0, clip, width=0.46\textwidth]{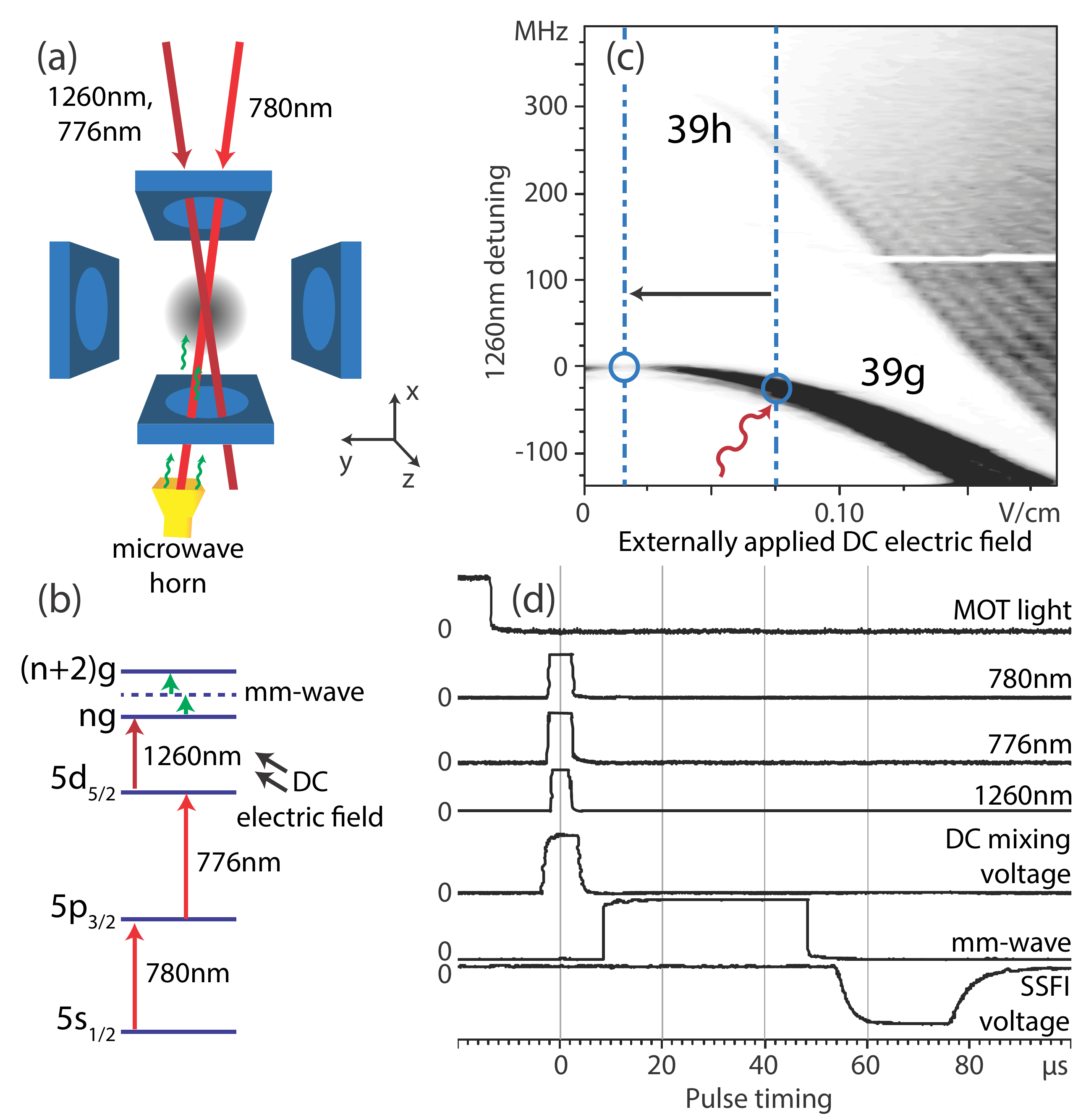}
  \caption{(Color online) Outline of the experiment. (a) Electrode, laser, and microwave-horn configuration in the experimental setup. The two \textit{z}-electrodes present in the experiment are not pictured. (b) Level diagram of excitation scheme. (c) Experimental Stark map demonstrating the preparation of $39g$ population. Black indicates high ($>$ 40) Rydberg counts per detection cycle. The preparation consists of optical excitation of a DC-field-perturbed $ng$ state (red squiggly line) and subsequent ramping of the DC field to zero (black arrow). (d) Partial timing diagram of experiment. Full experimental period is 5 ms and includes MOT loading time, not pictured here.}
  \label{figure 1}
\end{figure}

The experimental setup is shown in Fig. \ref{figure 1}(a), the energy level diagram in Fig. \ref{figure 1}(b), and the timing sequence of the experiment in Fig. \ref{figure 1}(d). Atoms are laser-cooled and trapped in a magneto-optical trap (MOT). During each experimental cycle, we prepare atoms in an initial Rydberg $ng$ state via an on-resonant three-stage optical excitation under simultaneous application of a perturbative DC electric field. The weak DC field admixes a small $nf$ character into the $ng$ state, allowing us to drive the $5d\rightarrow ng$ optical transition (see Fig. \ref{figure 1}(b)) in first order of the optical field. As it can be seen from the experimental Stark map shown in Fig. \ref{figure 1}(c), we observe significant population in the initial Rydberg $ng$ state, well-isolated from the neighboring $nh$ state and the hydrogenic manifold. After state preparation, the perturbative DC electric field is adiabatically lowered (see Fig. \ref{figure 1}(d)) to the pre-determined zero-field value, thereby producing a sample of $ng$-state atoms. Next, a rectangular microwave pulse is applied for  $\tau$ = 40 $\mu$s, driving the $ng\rightarrow(n+2)g$ transition. We scan the microwave frequency across resonance and detect the population in the target state $(n+2)g$ via state-selective field ionization (SSFI) \cite{gall}. \par

In our microwave system, microwaves are first generated in a synthesizer (Agilent N5183A). Next, they are frequency-quadrupled in an active frequency multiplier (Norden Millimeter N14-4680). The frequency-quadrupled output power can be varied over a limited range by adjusting the input power supplied by the synthesizer. While the input-to-output power relation is highly nonlinear~\cite{Moore2019a}, it nevertheless allows us to continuously vary the intensity at the location of the atoms over a range that allows us to observe the progression of the spectral lines from being indiscernible from the noise floor to being severely broadened, or until the output power of the multiplier saturates. The sub-THz intensity sweep is important in quantifying the AC-Stark shifts of the transitions when evaluating systematic uncertainties. \par

To reduce systematic shifts, we have specifically chosen to probe $\Delta l = 0$, $\Delta j = 0$ transitions. Since these transitions have equal Land\'e g-factors in the lower and upper states, for $\Delta {m_j}=0$ (our case), there is no line broadening due to the quadrupole field of the MOT or other external magnetic fields. Any $ \Delta {m_j} \ne 0$ transitions are broadened by the MOT magnetic field over hundreds of kHz and are not recognizable in our spectra. \par

We measure the frequencies of the $ng\rightarrow(n+2)g$ two-photon transitions for four choices of $n$. For each case, we take six data series. Five of the series are to evaluate systematics, as outlined in the following sections. The sixth data series is longer and is collected under conditions with minimized systematic uncertainty. The transition frequencies are extracted using two alternate methods that we refer to as Methods A and B. In Method A, we obtain the four transition frequencies from the long data series with minimized systematics. Method A produces a measurement result with low statistical uncertainty in the range of only a few 100~Hz. In Method B, we obtain the transition frequencies from maps of the transitions against microwave frequency and power. The data for Method B are taken from the measurement series that are collected to evaluate the systematic shifts and uncertainties from the AC Stark effect in both methods. It should be noted that Method B has a higher statistical uncertainty due to fewer measurements taken. However, this method provides the assurance that there are no AC shifts and mean-field shifts due to Rydberg-atom collisions. 

\section{Uncertainty analysis}
\label{sec:unc}

\begin{table}[t]
\caption{\label{tab:table1} Summary of shifts added to the measured frequencies and their uncertainties in kHz (see text for the details). DC$_i$ stands for the DC Stark shift in $i$-direction, and AC for AC Stark shift due to the drive microwave field. The clock shifts reflect corrections and uncertainties from the 10-MHz references used for the microwave synthesizer.}
\begin{ruledtabular}
\begin{tabular}{l | l | l | l | l}
    Shift(kHz) & $38g\rightarrow40g$ & $39g\rightarrow41g$ & $40g\rightarrow42g$ & $41g\rightarrow43g$\\ [3pt]
    \hline
    DC$_z$ & $0.1\pm 14$ & $0.01\pm4.7$ & $0.08\pm4.4$ & $0.02\pm4.1$\\ [3pt]
    DC$_x$ & $0.01\pm2.5$ & $0\pm2.6$ & $0.003\pm0.77$ & 
    $0.46\pm0.92$\\ [3pt]
    DC$_y$ & $0\pm1.6$ & $0.01\pm1.1$ & $0\pm1.4$ & $0\pm1.8$\\ [3pt]
    AC & $0\pm0.66$ & $0\pm0.62$ & $0\pm0.19$ & $0\pm0.37$ \\ [3pt]
    Clock & $0\pm0.011$ & $0\pm0.010$ & $4.78 \pm 0.19$ & $0\pm0.009$\\ [3pt]
    \hline
    Statistical & $\pm$0.28 & $\pm$0.18 & $\pm$0.22 & $\pm$0.30 \\ [3pt]
    \hline
    \textbf{Total} & \textbf{$0.11\pm14.3$} & \textbf{\textbf{$0.02\pm5.5$}} & \textbf{\textbf{$4.86\pm4.7$}} & \textbf{\textbf{$0.48\pm4.6$}} \\ [3pt]
\end{tabular}
\end{ruledtabular}
\end{table}

A careful analysis of statistical and, especially, systematic uncertainties is a critical component of the work. A summary of frequency corrections of our four measurements is given in Table \ref{tab:table1}. The uncertainties are given to two significant digits, so as to not lose information in the calculation of the corrected transition frequencies and their uncertainties (see Table \ref{tab:table2} in Sec.~\ref{sec:res}) \cite{taylor}.\par

One can see that systematic uncertainty from residual DC Stark shifts represents the largest source of uncertainty, followed by AC Stark shifts and statistical uncertainties. While the calibration offset of one of the employed 10-MHz reference sources is important, the resulting calibration uncertainty for each transition represents the smallest systematic effect. In the following sections, the procedure of the uncertainty evaluation is discussed in detail. 

\subsection{Statistical uncertainty and microwave-frequency calibration}
\label{subsec:statunc}

\begin{figure}[t]
 \centering
  \includegraphics[width=0.48\textwidth]{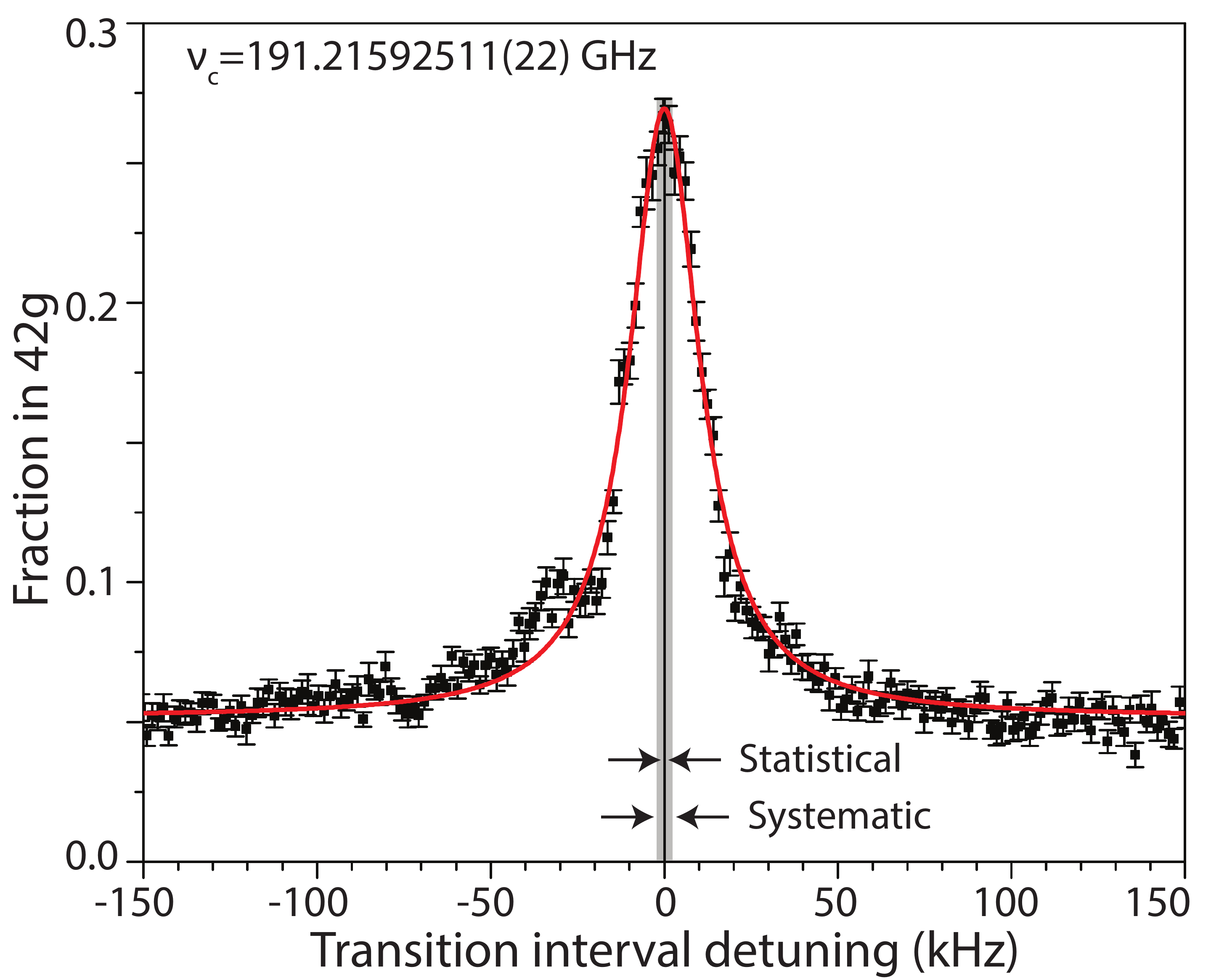}
  \caption{(Color online) Two-photon microwave spectrum of the $40g\rightarrow42g$ transition obtained with Method A explained 
  in Sec.~\ref{sec:methods}. Black squares, fraction of atoms detected in $42g$, detuned from $\nu_c$ (black line). The statistical uncertainty on $\nu_c$ is less than the width of the black vertical line. Uncertainty on the corrected interval frequency (gray region), listed in Table \ref{tab:table2}, reflects systematic uncertainty. Error bars on data points are SEMs. Red curve is a Lorentzian fit. In this Figure, the systematic corrections exhibited in Table \ref{tab:table1} have not yet been applied to $\nu_c$.}
  \label{figure 2}
\end{figure}

Averaging the long data series obtained after evaluation and minimization of systematics, we observe spectral peaks that approximate Fourier-limited $\mathrm{sinc^2}$-functions centered at frequencies $\nu_c$ (Fig. \ref{figure 2}). For our microwave interaction time $\tau$, the expected Fourier-sideband zeros at $m\times25$ kHz ($m$ is a nonzero integer) coincide with local minima observed in the spectrum. The sidebands are not well-resolved, indicating some inhomogeneous broadening and decoherence. The former arises from the presence of both $j=7/2$ and $j=9/2$ fine-structure transitions with $\Delta j =0$, $\Delta m_j=0$, which are separated by about 20~kHz. Here we assume that we drive a statistical mix of these transitions, centered at the fine-structure-free transition frequency. Comparatively small amounts of decoherence and inhomogeneous broadening may result from spontaneous Rydberg-atom decay, decay driven by black-body radiation, and residual quasi-static electric-field noise. Since the transitions are magnetic-field-insensitive, magnetic-field noise does not contribute to the broadening. Microwave noise and variations of the atom-field interaction time also do not contribute to inhomogeneous broadening or decoherence. \par 

Since a sideband-averaged $\mathrm{sinc^2}$ function approaches a Lorentzian, we perform each peak fit using a Lorentzian function. In Method A, in which we obtain the transition frequencies from an average over the long data series, we achieve a statistical uncertainty of the line centers a few 100~Hz. The statistical uncertainties are explicitly listed in Table~\ref{tab:table1}. Since the frequencies $\nu_{n_1, n_2}$ of the transition intervals (as defined in Eq.~\ref{eq:2}) exceed 0.1~THz, the statistical uncertainty amounts to a relative frequency uncertainty of the atomic transition frequencies of $\sim10^{-9}$. \par

At this level of precision, it is necessary to lock the internal crystal oscillator of the microwave synthesizer to an atomic reference. For three of the four cases of $ng$-states we have studied,  we have used a factory-calibrated, external atomic clock (SRS 725) with a relative uncertainty of $\pm5\times10^{-11}$. The absolute instrument uncertainty for the measured $ng\rightarrow(n+2)g$ frequency interval is $\approx10$ Hz. This instrument uncertainty is well below systematic uncertainties.\par

For the $40g\rightarrow42g$ frequency-interval measurement, we used a different atomic clock (DATUM LPRO) because the SRS 725 was not available. The LPRO clock had an unknown calibration due to aging. We have determined the LPRO clock shift by beating the LPRO clock with the calibrated SRS 725 used in the data sets for $n$ = 38, 39, and 41. We determined that the LPRO runs faster by a relative amount of $2.53\times10^{-8}$, showing that a correction accounting for the LPRO's clock shift was important. The frequency correction applied to the $40g\rightarrow42g$ measurement that results from the LPRO clock shift is explicitly exhibited in Table \ref{tab:table1}.

\subsection{DC Stark shifts}
\label{subsec:dc}

As seen in Table \ref{tab:table1}, the DC Stark shift, which scales as the square of the electric field, $E$, is the leading systematic uncertainty. As the electric-field components $E_i$ ($i = x, y, z$) add up in quadrature to a net field $E$, it is important to control all three components of the field. Our setup is designed such that this is possible. In order to minimize the shifts due to static electric fields, we follow the standard procedure \cite{lee, Ramos2019}. The field zeros are determined by varying a field direction $E_i$ ($i = x, y, z$) while holding the other directions fixed until a minimum shift of the transition frequency $\nu$ is observed. This is determined by measuring $\nu$ as a function of the DC tuning voltage that corresponds to field direction $i$ and fitting the result to a parabola. The uncertainty in the parabolic fit determines the uncertainty in the residual DC Stark shift contributed by the field direction $i$. The procedure is iteratively performed for all directions. The corrections and uncertainties derived from this procedure are listed in Table \ref{tab:table1} (the listed uncertainties include the noise of the tuning voltage sources). The uncertainty corresponding to the DC Stark shift in the $z$-direction is the dominant systematic. We attribute this to the fact that the $z$-direction field is applied via a high-voltage amplifier that is needed for SSFI. While the amplifier noise was mostly eliminated with a filter circuit, it was still noticeable.

\subsection{Rydberg-Rydberg interactions}
\label{subsec:rri}

\begin{figure}[t]
 \centering
  \includegraphics[width=0.45\textwidth]{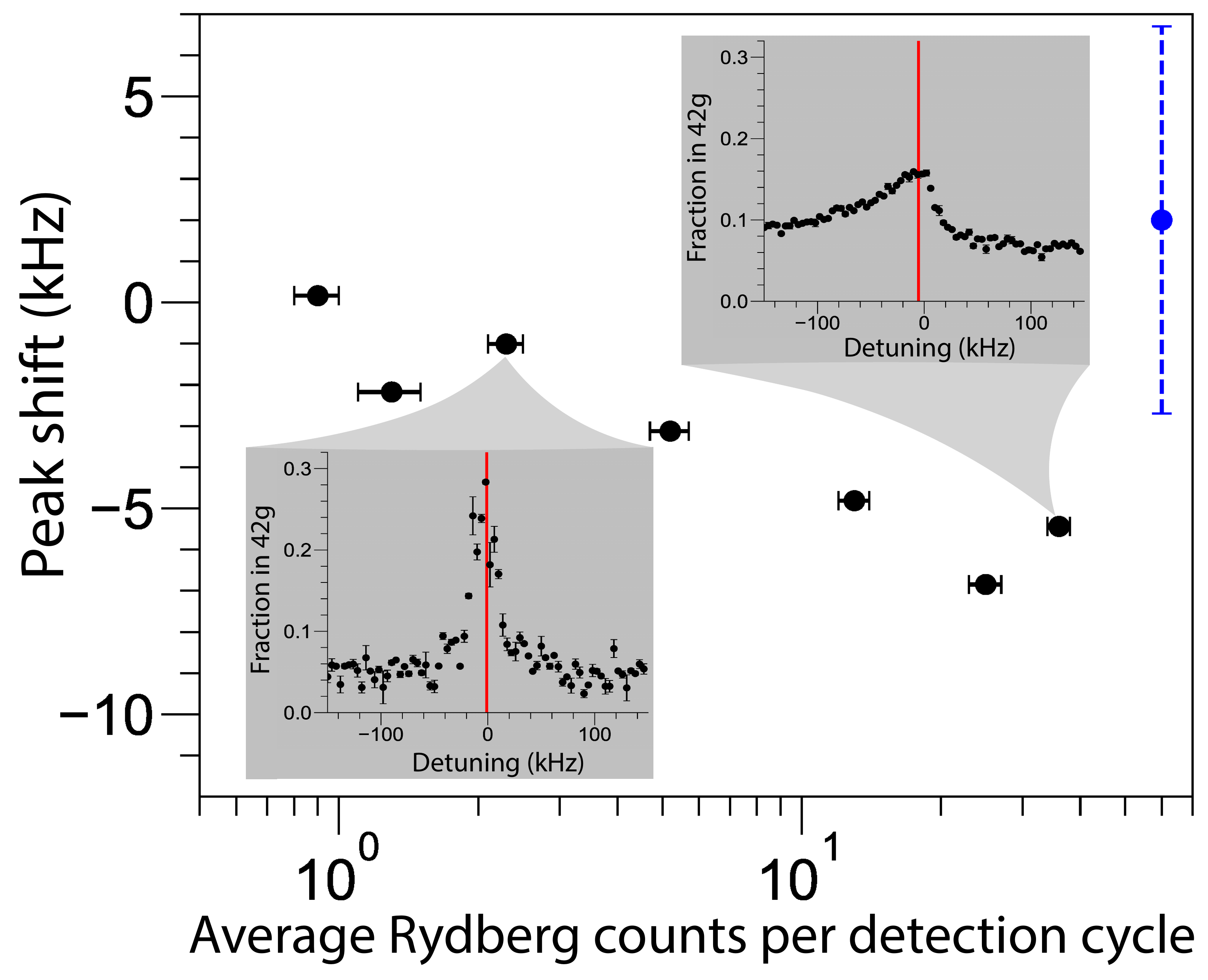}
  \caption{(Color online) Dependence of the $40g\rightarrow42g$ transition on Rydberg-atom density. This analysis is representative of the procedure used for all four measured transitions. Data points (black circles) show peak shift relative to the frequency values given in Table \ref{tab:table2} versus average Rydberg-atom counts per detection cycle. The peak shifts are obtained from local parabolic fits around the peaks in the spectra. The standard errors of the peak shifts are negligible. For comparison, in the upper-right corner we have added an artificial data point with an error bar that shows the systematic uncertainty provided in Table~\ref{tab:table1}. Horizontal error bars are the standard deviations of the counts. The insets display two measured spectra and peak positions (red vertical lines) extracted from the parabolic fits.}
  \label{figure 3}
\end{figure}

The effect of Rydberg-Rydberg interactions leads to asymmetric line broadening and mean-field line shifts if the number of Rydberg counts per sample is chosen too high. Here, we have identified a maximum number of Rydberg atoms that may be excited per detection cycle without inducing collisional shifts affecting the result, thereby optimizing the signal/noise ratio while avoiding collisional shifts. In Fig.~\ref{figure 3}, we show the measured peak position versus average total detected Rydberg counts per cycle for the $40g\rightarrow42g$ transition. The peak positions are obtained from parabolic fits over frequency intervals that cover the peaks in the spectra. Two of the spectra are shown in the insets of the Fig.~\ref{figure 3}. Although an increase in detected Rydberg counts generally improves the signal/noise ratio, Rydberg-Rydberg interactions are seen to cause a red shift of the detected transition frequency, as well as asymmetric broadening (compare insets in Fig.~\ref{figure 3}). The asymmetric lineshape observed at high Rydberg counts (upper-right inset in Fig. \ref{figure 3}) is attributed to details in the van der Waals interactions between Rydberg atoms, which we elaborate on in the next paragraph. For each transition, we observe that below about ten detected Rydberg counts per sample the peak position and linewidth coincide with their interaction-free values with an accuracy considerably better than the systematic uncertainty (see error bar on the right in Fig.~\ref{figure 3}). We therefore do not include Rydberg-Rydberg interactions in the analysis of shifts and uncertainties outlined in Table \ref{tab:table1}.\par

The long data series used in Method A of our analysis are taken in the regime of less than ten counts per sample. In this regime, the peaks in the spectra are fitted well with Lorentzian functions (see Fig.~\ref{figure 2} for one case), yielding the atomic transition frequencies without collisional shifts. 

\begin{figure*}[ht!]
 \centering
  \includegraphics[width=1\textwidth]{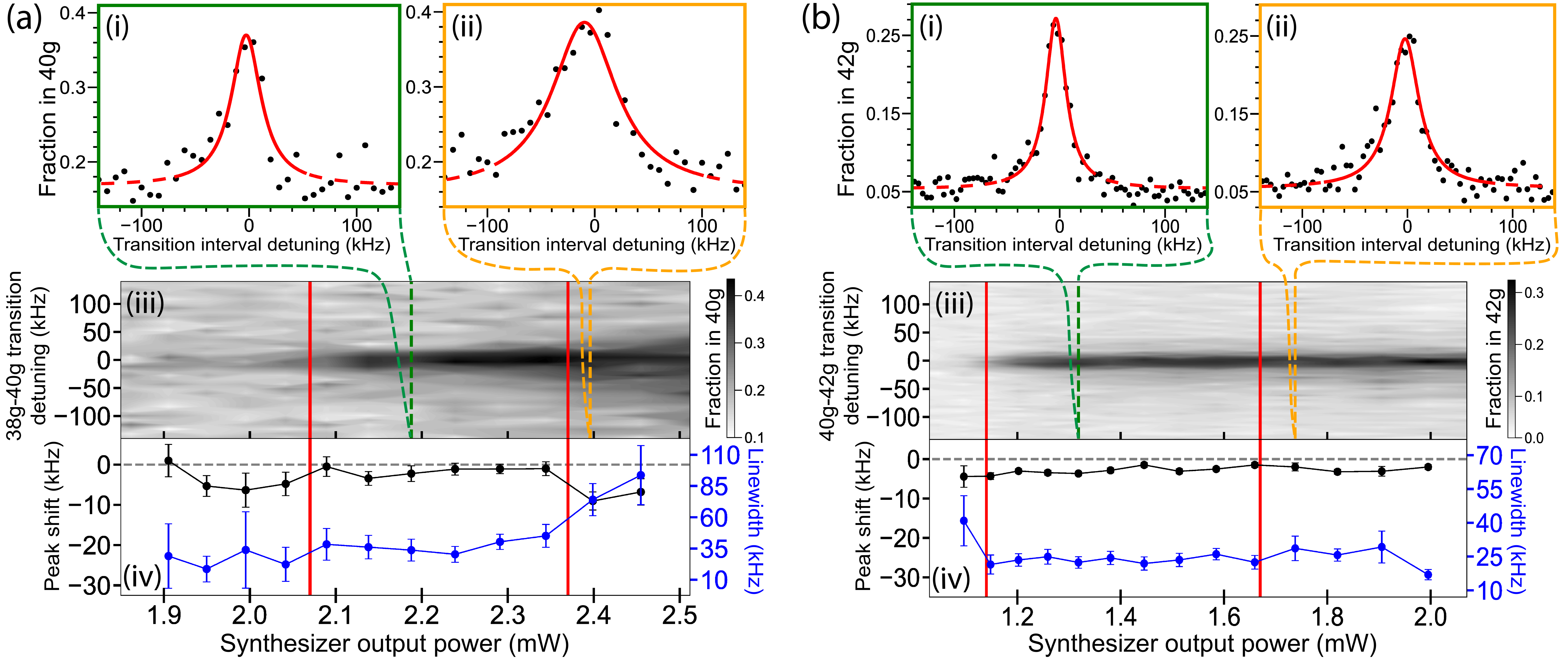}
  \caption{(Color online) AC Stark shift analysis for the $38g\rightarrow40g$ (a) and $40g\rightarrow42g$ (b) transitions. (i) Examples of spectra that are included in the determination of the transition frequency. (ii) Sample spectra that are asymmetrically broadened and are not included. (iii) Spectral signal on linear gray scales vs transition detuning (vertical) and microwave power supplied into quadrupler. (iv) Center frequency and spectral linewidth vs microwave power.  In (i)-(ii), the solid red lines show Lorentzian fit functions over the spectral ranges used for the fits, whereas the dashed red lines show extrapolations of the fits over the entire widths of the plots. In (iii)-(iv), the vertical solid red lines delimit the power range used for analysis of the transition frequencies. In this Figure, the systematic corrections exhibited in Table \ref{tab:table1} have not been applied yet, resulting in a small offset of the center frequencies from zero detuning.}
  \label{figure 4}
\end{figure*}

For a theoretical assessment of the shifts expected due to pairwise interactions, we have calculated the adiabatic molecular potentials of Rydberg-atom pairs using the methods from~\cite{Schwettmann2006, Deiglmayr2014, Sassm2016, Han2018, Han2019}.  We considered  the molecular configurations that both interacting atoms are in \(ng\), both atoms are in \((n+2)g\), and one is in \(ng\) and the other in \((n+2)g\), and for projections of angular momentum onto the internuclear axis ranging from $M=0$ to $\vert M \vert =9$. All adiabatic interaction potentials exhibit a clean van der Waals dependence $\propto R^{-6}$, at internuclear distances $R \gtrsim 3~\mu$m. There are about twice as many attractive than there are repulsive potentials, and the density of potentials is greatest near zero. The shifts range between maximal values of about  $- h \times 7.5 \times 10^{-44} (n_1 \, n_2)^{5.5} / R^6$~Hz~m$^6$ and $h \times 5.0 \times 10^{-44} (n_1 \, n_2)^{5.5} / R^6$~Hz~m$^6$, with $n_1$ and $n_2$ denoting the principal quantum numbers of the involved $g$ states (here, $n_2 = n_1 + 2$). Since the excitation volume is estimated at about 1~mm$^3$ and the detection efficiency at about $30\%$, at the highest count numbers in Fig.~\ref{figure 3}, the average internuclear separation in a frozen gas would be on the order of 100~$\mu$m, leading to van der Waals shifts in the in sub-Hz range. The van-der-Waals forces and accelerations at the average internuclear separation are too small to cause any substantial movement of the atoms during the interaction time $\tau=40~\mu$s. On the other hand, during the interaction time $\tau$, thermal motion is estimated to bring a large fraction of Rydberg atoms into proximity with another Rydberg atom (distances in the range of 10$~\mu$m). At this distance, atom pairs come within range of substantial van-der-Waals forces (which scale as $R^{-7}$). The force may be attractive or repulsive, depending on which adiabatic potential the pair evolves. Because attractive forces are conducive to closer collisions than repulsive forces, and because Rb $ng$ atom pairs at a given $R$ experience negative level and transition shifts generally larger in magnitude than the positive shifts, it is expected that the interactions should lead to asymmetric spectra biased towards negative detunings, as observed in the higher-density spectrum in Fig.~\ref{figure 3}.\par

The dynamics may also involve a fraction of spectator Rydberg atoms in $nf$ states in the sample, which could become populated by black-body-radiation-induced transitions or minor non-adiabaticity in the excitation sequence shown in Fig.~\ref{figure 1}. These would exert electric-dipole forces onto the $g$-type atoms. Electric-dipole forces are stronger than van der Waals forces, and the above outlined mechanism may become even more visible. \par

Spectra in interacting Rydberg-atom systems have been observed and discussed in earlier work, including~\cite{Han2009, Beguin2013, Reinhard2008, Reinhard2007}. These previous studies were performed with low-angular-momentum Rydberg atoms and have shown pronounced asymmetric line shapes resembling the upper-right inset in Fig.~\ref{figure 3}. While a detailed examination of Rydberg-atom interactions in dense samples of  $ng$ Rydberg atoms may be an interesting topic in future work, the present experiment is conducted at low densities such that the transitions of interest are not significantly affected by these interactions.

\subsection{AC Stark shifts}
\label{subsec:ac}

We have timed the excitation sequence such that optical light is not present during the measurement interval, eliminating optical AC Stark and ponderomotive shifts (see Fig. \ref{figure 1}(d)). However, the shifts due to the microwave field and their interrelation with Rydberg-atom collisions must be estimated. Typically, an AC shift correction may be performed by measuring line frequency vs microwave power and extrapolating the line shift to zero power \cite{lee, berl}. This is not possible here due to the nonlinearity of the frequency multiplier, which has been calibrated by us~\cite{Moore2019a} using an atom-based RF field measurement method~\cite{Sedlacek2012, Holloway2014}. The multiplier output power approximately scales as the ninth power of the input power and saturates at about 2~mW output power. In the present experiment, the useful variation range of the input power is $\lesssim 1.5$~dB, resulting in a range of about $\lesssim 15$~dB in output power, corresponding to a range of about a factor of thirty of the two-photon Rabi frequeny. In view of these characteristics, we measure the spectra versus signal generator power injected into the quadrupler for each of the four transitions studied. For the determination of the transition frequencies, we then use only injected microwave powers within ranges as indicated by the vertical red lines in Fig.~\ref{figure 4} (a)-(b) (iii)-(iv)). Within these injected power ranges, the quadrupler output power monotonically increases, the spectra have a clearly visible peak, have no significant asymmetry, and yield transition-frequency results that are stable against modest variations of the power. In the following, we justify this procedure. \par 

At low injected microwave power, the signal is barely visible above the noise floor, making the peak centers hard to determine. As the power increases, the lines become more visible. In a few cases, there is a hint of a positive  AC shift that slightly exceeds the statistical uncertainty. This is seen in the trends of the peak shifts plotted Fig.~\ref{figure 4} (iv). A small positive AC shift of the transition frequency is consistent with our calculations (see below), which show that both the $ng$ and $(n+2)g$ states have negative AC shifts. These shifts mostly cancel in the transition frequency. Since the lower state shifts more than the upper state, the transition frequency exhibits a slight net increase (positive AC shift). \par

At the highest powers available, we observe an asymmetric line broadening similar to that caused by Rydberg-Rydberg interactions (see Fig.~\ref{figure 3}, upper-right inset). We interpret this behavior as follows. As discussed in Sec.~\ref{subsec:rri}, a fraction of the atoms may come close to each other in the course of the atom-field interaction duration $\tau = 40~\mu$s. At time instances of close approach, the $ng$ levels are shifted according to the van-der-Waals shift equations provided. The transition frequencies exhibit negative or positive shifts, with a preponderance of negative shifts. At low microwave power, the RF Rabi frequencies are too small to effectively drive detuned transitions during close encounters, because the transition-frequency shifts are rapidly changing during these events. As the RF Rabi frequency increases, transitions during close encounters become noticeable, causing negative- and positive-shifted wings at high power in Fig.~\ref{figure 4}. In accordance with the van der Waals interaction potentials, the negative-shifted wings are more significant. Even if the wings extend to 100~kHz or more, the mean-field shift of the line centers still amounts to less than about 10~kHz. This is because the RF transitions are heavily dominated by time segments during which Rydberg-atom pairs are at large relative distances from each other, where there is no collision-induced shift and the transition frequency is time-independent. Due to the low average density of the Rydberg-atom sample, the shift-free time intervals cover most of the $\tau = 40~\mu$s atom-field interaction time, Hence, the peak shift is much smaller than the overall width of the lopsided spectra.\par

To avoid systematic effects from collisions, in our analysis of the transition frequencies we only process spectra within the vertical red cursors in Fig.~\ref{figure 4} (iii) and (iv), where the RF Rabi frequency is too small to significantly drive transitions during short-range encounters between Rydberg atoms. To ensure that any remaining collision-induced asymmetry as well as asymmetry in the broadened Fourier sidebands of the peaks do not affect the fit results for the line centers, we restrict the frequency range used for fitting of the peak centers to a narrow region around the maximum value of Rydberg counts (solid red curves in Fig.~\ref{figure 4} (a)-(b) (i)-(ii)). The fits are then extrapolated over the full range (dashed red curves). \par

For a theoretical estimate of the AC shift, we have calculated AC shifts and two-photon Rabi frequencies for $m_j=-9/2$ to 9/2. Both the AC shifts of the transition frequency and the Rabi frequencies scale as the square of the RF electric field, $E_{RF}$. For the transitions studied in our work, the AC shifts of the transition frequencies are typically $10\%$ of the Rabi frequencies. The AC shift scales approximately as $n^{6.6}$, and the Rabi frequencies as $n^8$. For example, for $\vert m_j \vert =0.5$ and $n=40$ the Rabi frequency in rad/s is $2 \pi \times 3.52~$kHz/(V/m)$^2 \times E_{RF}^2$, while the AC shift of the transition frequency, in Hz, is 209~Hz/(V/m)$^2 \times E_{RF}^2$. Close to saturation of the transitions, the upper-state population becomes maximal. Saturation of the upper-state population occurs near the upper bound of injected microwave power of the spectra we process. For our atom-field interaction time of $\tau = 40~\mu$s, the system approaches upper-state population saturation when the Rabi frequency becomes on the order of $2 \pi \times 10$~kHz. This value corresponds with RF electric fields of about 1.5~V/m and AC shifts of the transition frequency of about 500~Hz, corroborating the finding that the lines do not exhibit a significant and quantifiable AC shift. \par

Further verification of the Rabi frequency in our spectra can be obtained from the power-broadening behavior seen Fig.~\ref{figure 4} (iv). The peak linewidths within the processed ranges of the injected microwave powers are only slightly larger than the Fourier-limited linewidth, $\gamma_F=0.89/\tau=22.3$~kHz. Hence, the processed spectra exhibit no or only mild saturation broadening, allowing an upper-bound estimate of the microwave Rabi frequency in these spectra of $\sim 2 \pi \times 25~$kHz/2, which is $\sim 2 \pi \times 10$~kHz. This value agrees with the estimate in the previous paragraph, also leading to the conclusion that the lines do not exhibit a significant and quantifiable AC shift. \par

In line with the above findings, in Table~\ref{tab:table1} we report a zero AC shift. The reported uncertainties of the AC shifts are given by the uncertainty in the weighted average of the center frequencies of the spectra, taken over the selected ranges of injected microwave power (see Fig.~\ref{figure 4}).

\maketitle
\section{Results} 
\label{sec:res}

\subsection{Method A}
\label{subsec:resA}

In Method A, we derive the transition frequencies from  Lorentzian fits to the sixth (long) data series (see Sec.~\ref{sec:methods}), one case of which is shown in Fig.~\ref{figure 2}. The systematic corrections given in the last row in Table~\ref{tab:table1} are then applied to the measured transition frequencies. The results of Method A for the transition frequencies are summarized in Table~\ref{tab:table2}. The uncertainty values in Table~\ref{tab:table2} are identical with the total uncertainties provided in Table~\ref{tab:table1}. \par

\begin{figure}[b!]
 \centering
  \includegraphics[trim=3 0 0 10, clip,width=0.45\textwidth]{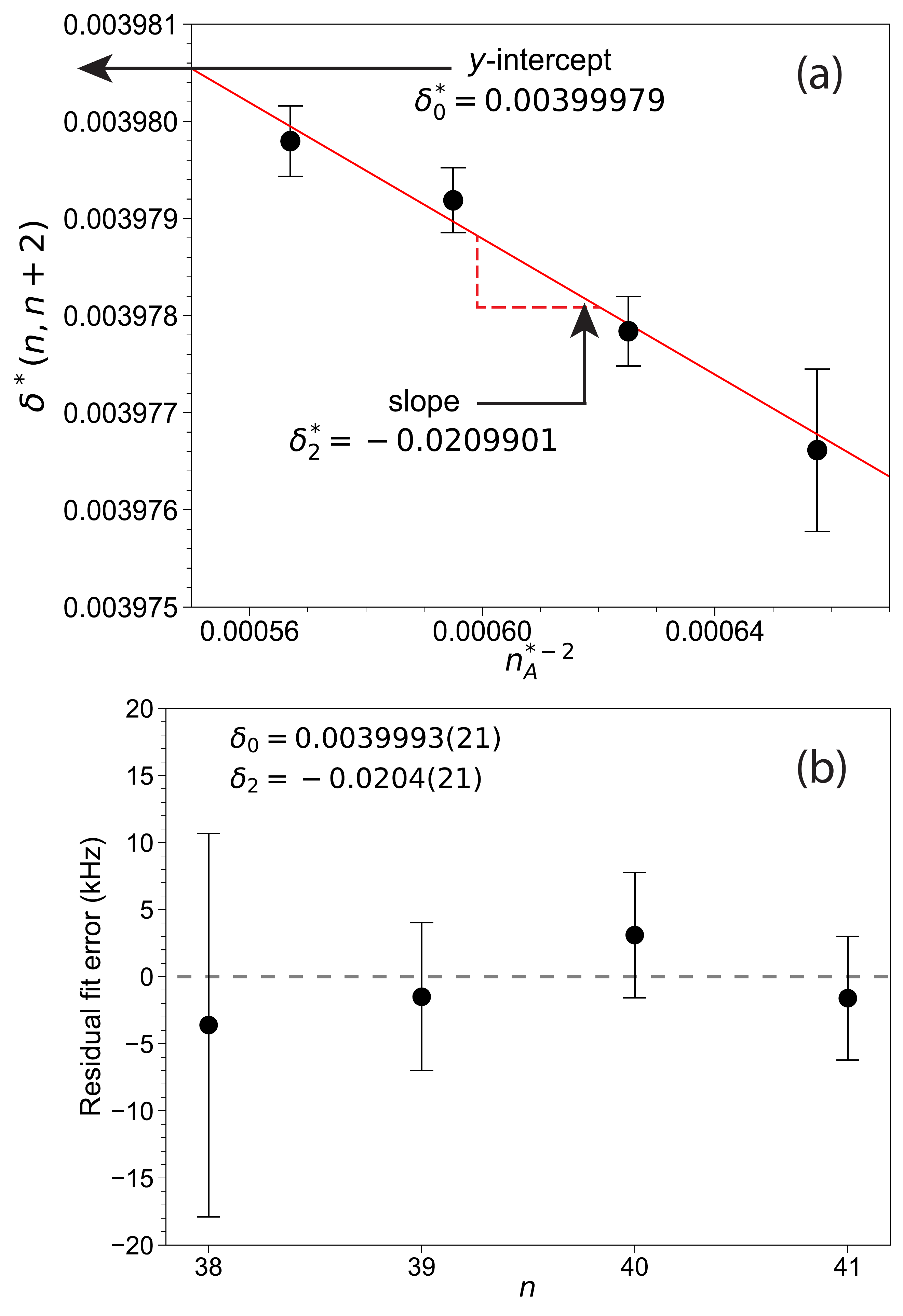}
  \caption{(Color online) Quantum defect determination from the 
  data in Table \ref{tab:table2}. (a) Determination of the seed values $\delta_0^*$ and $\delta_2^*$ for fitting. Black circles are data. Error bars are errors propagated from the frequency-interval results in Table \ref{tab:table2}. Red line is a weighted linear fit to data. (b) Determination of $\delta_0$ and $\delta_2$ from a nonlinear least-squares fit using the model in Eq.~(\ref{eq:2}), initialized with $\delta_0^*$ and $\delta_2^*$. The plot shows the residuals of the fit relative to the frequency-interval results in Table \ref{tab:table2}, with error bars propagated from Table \ref{tab:table2}.}
  \label{figure 5}
\end{figure}

To obtain values for the $ng$-series $\delta_0$ and $\delta_2$ quantum defects, we follow the procedure of \cite{li}. First, we obtain an average quantum defect $\delta^*(n,n+2)$ for the pair of microwave-coupled levels using

\begin{equation}
\label{eq:3}
\nu_{n} = R_{Rb} \ c \ \left( 
  \frac{1}{(n - \delta^*(n,n+2))^2} 
- \frac{1}{(n + 2 - \delta^*(n,n+2))^2} 
\right), 
\end{equation}

\noindent In Table \ref{tab:table2}, we list the values for $\delta^*(n,n+2)$ and their uncertainties together with the transition frequencies $\nu_{n, n+2}$ obtained with Method A.\par

Next, in Fig. \ref{figure 5}(a) we plot $\delta^*(n,n+2)$ versus $\bar{n}^{*-2}$, where the average $\bar{n}^* = n+1-\delta_0^G$. Here, $\delta_0^G$ is a previously determined value of $\delta_0$~\cite{lee}. Extrapolating to $\bar{n}^{*-2} = 0$, we obtain an initial estimate $\delta_0^* = 0.00399979$. For each of the four transitions studied, we then substitute $\delta(n) = \delta_0^* + \delta_2^*/(n-\delta_0^*)^2$ into Eq.~\ref{eq:2} and solve for $\delta_2^*$. Averaging the four results yields a preliminary value $\delta_2^* = -0.0209901$. It is noted that the initial seed values $\delta_0^*$ and $\delta_2^*$ do not affect the actual final fit result, shown next. \par

Using $\delta_0^*$ and $\delta_2^*$ as initial values for the two free parameters $\delta_0$ and $\delta_2$ in Eqs~\ref{eq:1} and~\ref{eq:2}, we perform a nonlinear least-squares fit to the transition-frequency values listed in Table \ref{tab:table2}, where $n$ is the independent variable. We use a Levenberg-Marquardt algorithm with assigning weights of the data points to $1/\sigma_i^2$ ($\sigma_i$ being a frequency uncertainty of the $i$-th data point).\par 

Our results for the $ng$-series quantum defects based on Method A are $\delta_0 = 0.0039993(21)$ and $\delta_2 = -0.0204(21)$. The uncertainties in these results include the propagation of the uncertainties listed in Table~\ref{tab:table1}. Systematic uncertainty of DC Stark shifts is the dominant contributor to the uncertainties of the quantum defects. The residuals in the transition frequencies that result from the fit, along with their uncertainties, are plotted in Fig. \ref{figure 5}(b).

\begin{table}[t]
\caption{\label{tab:table2} Summary of results for the transition 
frequencies $ng \rightarrow (n+2)g$ from Method A.}
\begin{ruledtabular}
\begin{tabular}{lcr}
$n$ & Transition frequency (GHz) & $\delta^*(n, n+2)$\\
\hline
38 & 222.199268(14) & 0.00397661(82)\\
39 & 205.9325351(55) & 0.00397784(36)\\
40 & 191.2159300(47) & 0.00397919(34)\\
41 & 177.8690737(46) & 0.00397980(36)\\
\end{tabular}
\end{ruledtabular}
\end{table}

\subsection{Method B}
\label{subsec:resB}

In our alternate Method B, we use the transition frequencies determined in the the course of the AC Stark shift analysis described in Sec.~\ref{subsec:ac} to extract the \(\delta_0\) and \(\delta_2\) quantum defects. The transition frequencies and corresponding values of $\delta^*(n,n+2)$ from Method B are displayed in Table~\ref{tab:table3}. The results according to Method B are $\delta_0 = 0.0039985(26)$ and $\delta_2 = -0.0197(29)$. Here, we use the same DC and clock shifts and uncertainties as listed in Table~\ref{tab:table1}. 

\begin{table}[h]
\caption{\label{tab:table3} Summary of results for the transition frequencies $ng \rightarrow (n+2)g$ from Method B.}
\begin{ruledtabular}
\begin{tabular}{lcr}
$n$ & Transition frequency (GHz) & $\delta^*(n, n+2)$\\
\hline
38 & 222.199266(14) & 0.00397652(82)\\
39 & 205.9325349(55) & 0.00397783(36)\\
40 & 191.2159319(47) & 0.00397933(33)\\
41 & 177.8690699(46) & 0.00397950(36)\\
\end{tabular}
\end{ruledtabular}
\end{table}

\subsection{Summary of Results}
\label{subsec:resfinal}
Foremost, we observe that the results of Method B are not statistically different from those of Method A, providing a consistency check of our analyses. By computing a weighted average of the values obtained using Methods A and B, we determine
\begin{eqnarray}
\delta_0 & = & 0.0039990(21) \nonumber \\
\delta_2 & = & -0.0202(21) \nonumber
\end{eqnarray} 
as our final results for the $g$-series quantum defects. Since the weights scale as the inverse-square of the uncertainties, the final result is dominated by Method A. In the final result we maintain the uncertainties from Method A, because the leading uncertainty, the systematic uncertainty due to the DC Stark effect, is the same in both cases, making the uncertainties of the methods A and B largely dependent. We also note that more data were collected in Method A than in Method B.

\section{Discussion}
\label{sec:disc}

Although our results are consistent with a recent measurement presented in \cite{lee} and at least one order of magnitude more precise, there is a relative difference in $\delta_0$ of $\sim 10^{-5}$ and in $\delta_2$ of $\sim10^{-2}$ compared with the most recent report in \cite{berl}. While a different atom-field interaction time is utilized ($40 ~\mu$s) in our present experiment, and our results benefit from the stray electric field control in three directions, the aforementioned discrepancies remain to be resolved. \par

In order to extract dipolar, $\alpha_d$, and quadrupolar, $\alpha_q$, polarizabilities of the Rb$^+$ ionic core from measured transition frequencies, one must adapt the method of \cite{lee} to measurements of shifts of transition energies relative to their quantum-defect-free values, $\delta W$, by writing (in atomic units)

\begin{align}
\label{eq:4}
2 \, \delta W \frac{1}{\langle1/r^4\rangle_D}= \alpha_d + \alpha_q \frac{\langle1/r^6\rangle_D}{ \langle1/r^4\rangle_D}.
\end{align}

\noindent Here $\langle1/r^i\rangle_D = \langle1/r^i\rangle_{n,l} - \langle1/r^i\rangle_{n+2,l}$. As can be seen by examining the $\langle1/r^i\rangle_{n,l}$ functions in \cite{bethe, Drake1990}, the values of $\langle1/r^i\rangle_{n,l}$ depend much more on $l$ than on $n$. The $n$-dependence of $\delta^*(n, n+2)$, evident in Table \ref{tab:table2}, is not sufficient to allow for a determination of $\alpha_d$ and $\alpha_q$ because all states have the same $l$. The $\Delta l=0$ sub-THz method presented in this paper is, in principle, well-suited for a measurement of the $nh$-series and higher-$l$ quantum defects. Also, improved magnetic-field control, as employed in~\cite{Ramos2019} for measurements of Rydberg-atom hyperfine structure, may allow spectroscopy of optical-molasses-cooled Rydberg-atom transitions with different $l$ in the initial and target states. These modifications may allow a measurement of Rb$^+$ polarizabilities in the future.

\maketitle
\section{Conclusion} 
\label{sec:concl}
In summary, we have presented a measurement of the $g$-series quantum defects using two-photon microwave spectroscopy of laser-cooled Rb atoms and compared our results with~\cite{lee,berl}, noting a discrepancy with~\cite{berl}. While the probing of \(\Delta l=0\) transitions eliminates the Zeeman effect from external magnetic fields, careful control over stray electric fields in all three directions with in-vacuum electrodes has been found to be critical in reducing systematic uncertainty. As evident in Table~\ref{tab:table1}, our leading systematic arises from the noise in the voltage applied to the $z$-direction electrodes. Experimental improvement is possible through a re-design of the SSFI apparatus. Possible extensions of the work based on this and  other improvements have been discussed in Sec.~\ref{sec:disc}.

\maketitle
\section*{ACKNOWLEDGMENTS}
KRM acknowledges support from the University of Michigan Rackham Pre-Doctoral Fellowship. This work was supported by NSF Grant No.PHY1506093 and NASA Grant No. NNH13ZTT002N NRA.

\bibliography{references}
\bibliographystyle{apsrev}

\end{document}